\documentclass[final,5p,times,twocolumn]{elsarticle}

\usepackage{lineno,hyperref}
\modulolinenumbers[5]

\usepackage{numcompress}

\begin{document}

\begin{frontmatter}

\title{The Tangerine project: Development of high-resolution 65\,nm silicon MAPS}


\author{Håkan Wennlöf\corref{mycorrespondingauthor}}
\cortext[mycorrespondingauthor]{Corresponding author}
\ead{hakan.wennlof@desy.de}

\author{Ankur Chauhan}
\author[bonn]{Manuel Del Rio Viera}
\author{Doris Eckstein}
\author{Finn Feindt}
\author{Ingrid-Maria Gregor}
\author{Karsten Hansen}
\author{Lennart Huth}
\author[campinas]{Larissa Mendes}
\author{Budi Mulyanto}
\author[wuppertal]{Daniil Rastorguev}
\author{Christian Reckleben}
\author[bonn]{Sara Ruiz Daza}
\author{Paul Schütze}
\author[bonn]{Adriana Simancas}
\author{Simon Spannagel}
\author{Marcel Stanitzki}
\author{Anastasiia Velyka}
\author[bonn]{Gianpiero Vignola}
\address{Deutsches Elektronen-Synchrotron DESY, Notkestr. 85, 22607 Hamburg, Germany}

\fntext[bonn]{Also at University of Bonn, Germany}
\fntext[campinas]{Also at University of Campinas, Brazil}
\fntext[wuppertal]{Also at University of Wuppertal, Germany}


\begin{abstract}
The Tangerine project aims to develop new state-of-the-art high-precision silicon detectors. Part of the project has the goal of developing a monolithic active pixel sensor using a novel 65\,nm CMOS imaging process, with a small collection electrode. This is the first application of this process in particle physics, and it is of great interest as it allows for an increased logic density and reduced power consumption and material budget compared to other processes. The process is envisioned to be used in for example the next ALICE inner tracker upgrade, and in experiments at the electron-ion collider.

The initial goal of the three-year Tangerine project is to develop and test a sensor in a 65\,nm CMOS imaging process that can be used in test beam telescopes at DESY, providing excellent spatial resolution and high time resolution, and thus demonstrating the capabilities of the process. 
The project covers all aspects of sensor R\&D, from electronics and sensor design using simulations, to prototype test chip characterisation in labs and at test beams. The sensor design simulations are performed by using a powerful combination of detailed electric field simulations using technology computer-aided design and high-statistics Monte Carlo simulations using the Allpix$^2$ framework. 
A first prototype test chip in the process has been designed and produced, and successfully operated and tested both in labs and at test beams.

\end{abstract}


\end{frontmatter}


\section{Introduction}

The Tangerine project (Towards Next Generation Silicon Detectors) has the goal of developing and investigating particle detection sensors in new silicon technologies. In the presented part of the project, a monolithic active pixel sensor (MAPS) in a new 65\,nm CMOS imaging process is developed. In MAPS, the readout electronics and the sensitive volume are located in a single silicon chip, which differs from the commonly used hybrid sensors where the readout electronics and the sensitive volume are in two different chips bonded together. MAPS thus have a lower material budget, reduced complexity in terms of module assembly, and reduced production cost compared to hybrid sensors.
The new 65\,nm CMOS imaging process allows for a higher logic density in the sensor compared to previous MAPS produced in larger technology nodes, which makes smaller or more intelligent pixels possible and allows a lower overall power consumption. Sensors developed in this technology can thus have a wide range of applications, from future collider experiment detectors to astroparticle, photon science, and medical imaging detectors. The technology has only recently become available for particle physics applications however, and the Tangerine project aims to investigate its performance. The primary three-year goal of the project is to develop and test a sensor in a 65\,nm process that can be used to create a beam telescope at the DESY II Test Beam facility, with excellent spatial and temporal resolution. The performance goals of the new sensor is a position resolution below 3\,\textmu m, a time resolution finer than 10\,ns, and a very low material budget (corresponding to at most 50\,\textmu m of silicon) to reduce the effect of multiple Coulomb scattering. The sensor will also be designed to produce a per-pixel charge measurement, which can be used to improve the sensor position resolution.

Utilising a so-called quadruple-well technology, the sensor can have a small collection electrode while also having sophisticated full CMOS in-pixel electronics. This is beneficial as it reduces the sensor capacitance, which increases signal gain and reduces noise. The design also builds on from results of sensors using a 180\,nm CMOS imaging process containing sensor volume modifications that increase the size of the depleted region~\cite{modifiedProcess, ngapProcess}, with the goal of achieving a fully depleted MAPS sensor (i.e. DMAPS).

The Tangerine project covers all aspects of sensor development, from electronics engineering and the design of sensor geometry using simulations, to laboratory and test beam investigations of prototype test chips. At the time of writing, test structures from the first submission are being investigated in the lab and at test beams. The next submission will incorporate design changes motivated by the results of tests and simulations, and a larger pixel matrix. The submissions are made as part of an international collaboration of several institutes with an interest in 65\,nm MAPS developments.

\section{Sensor design and testing}

\subsection{Sensor design}

The main purpose of the first test chip designed at DESY is to investigate a newly designed fast charge-sensitive amplifier circuit. There are two amplifier variants available on the chip, each with a Krummenacher type feedback network for continuous reset and leakage current compensation~\cite{krummenacher}. The amplifier variants differ in their nominal gain. The chip also contains a $2 \times 2$ pixel matrix, with the possibility to read out the analogue signal of each pixel individually. The pixels in the matrix are square, with a side length of 16.3\,\textmu m. Most of the readout electronics are located outside of the pixels in this test chip.

Electronics and readout design for future sensor submissions is ongoing. The final sensor is intended to have a pixel matrix size of $256 \times 256$ pixels, and a pixel size as small as possible while containing the necessary readout electronics within the pixel. To keep the pixel size minimal, the shape may be changed from square to rectangular. A hexagonal pixel shape is also considered, to improve detection efficiency. Each pixel is intended to measure the signal amplitude via a time-over-threshold measurement.

\subsection{Lab and test beam investigations}

\begin{figure}[b]
\centering
    \includegraphics[width=0.45\textwidth, trim={0 0 12mm 13mm}, clip]{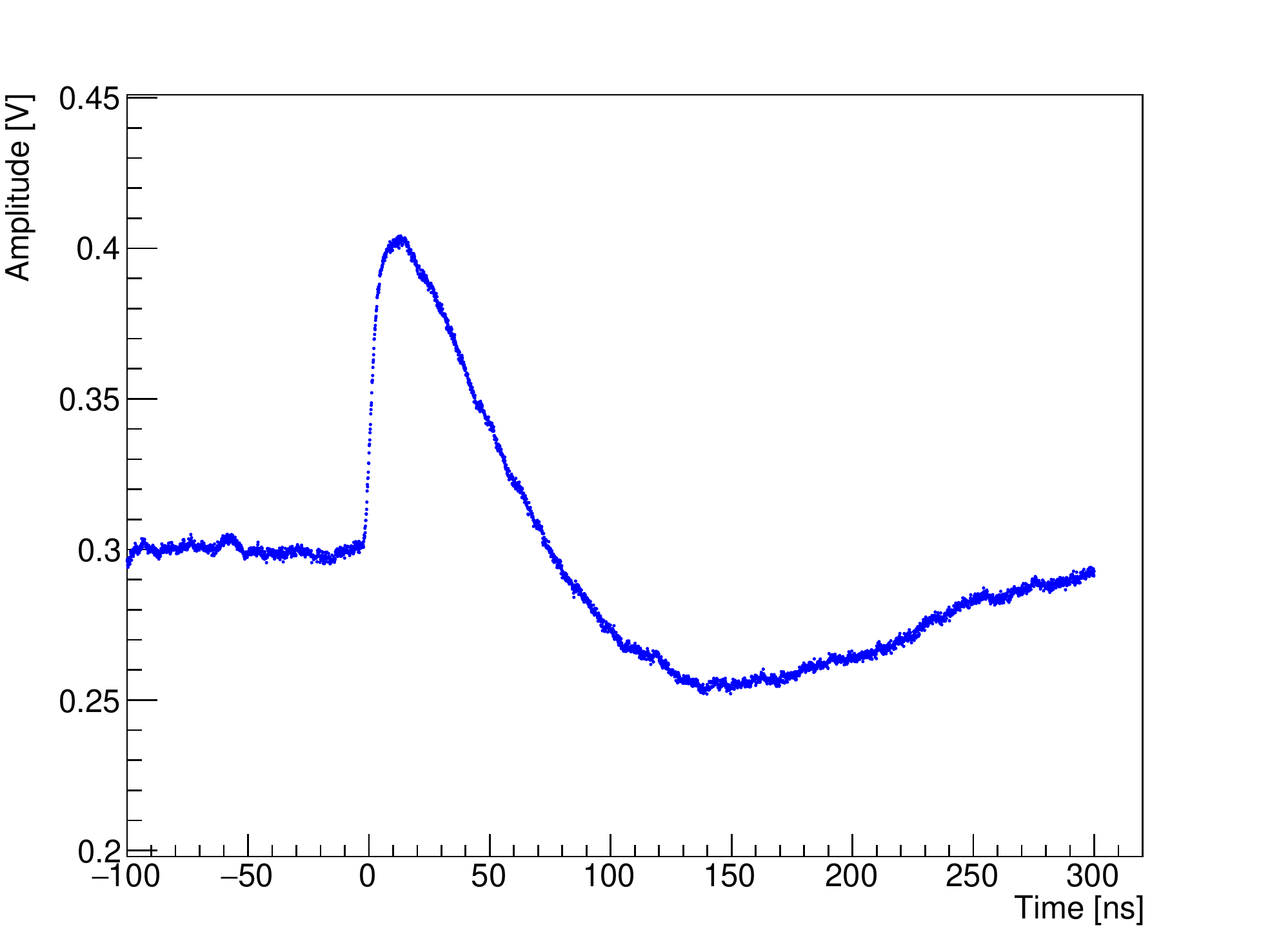}
    \caption{Single pixel hit waveform example, showing a hit from an x-ray produced by a $^{55}$Fe source.}
    \label{fig::waveformExample_1}
\end{figure}

The DESY test chip is characterised by tests performed in labs and at test beams. Lab tests are performed using radioactive sources and charge injection pulses, which makes it possible to extract signal waveforms from each pixel on the chip. By analysing the waveforms, the behaviour of the in-pixel charge-sensitive amplifiers can be studied. Figure~\ref{fig::waveformExample_1} shows an example waveform from a pixel hit by a $^{55}$Fe x-ray.
The pixel signal is constant (except for noise) before the x-ray hits, then rises rapidly to a maximum amplitude value before decaying back to the constant baseline via an undershoot.
Study of the waveforms can also give insight into how the sensor response is affected by different biasing and control current settings, and scans of different settings have been performed.

\begin{figure}[tbp]
\centering
    \includegraphics[width=0.42\textwidth, trim={0 0 12mm 13mm}, clip]{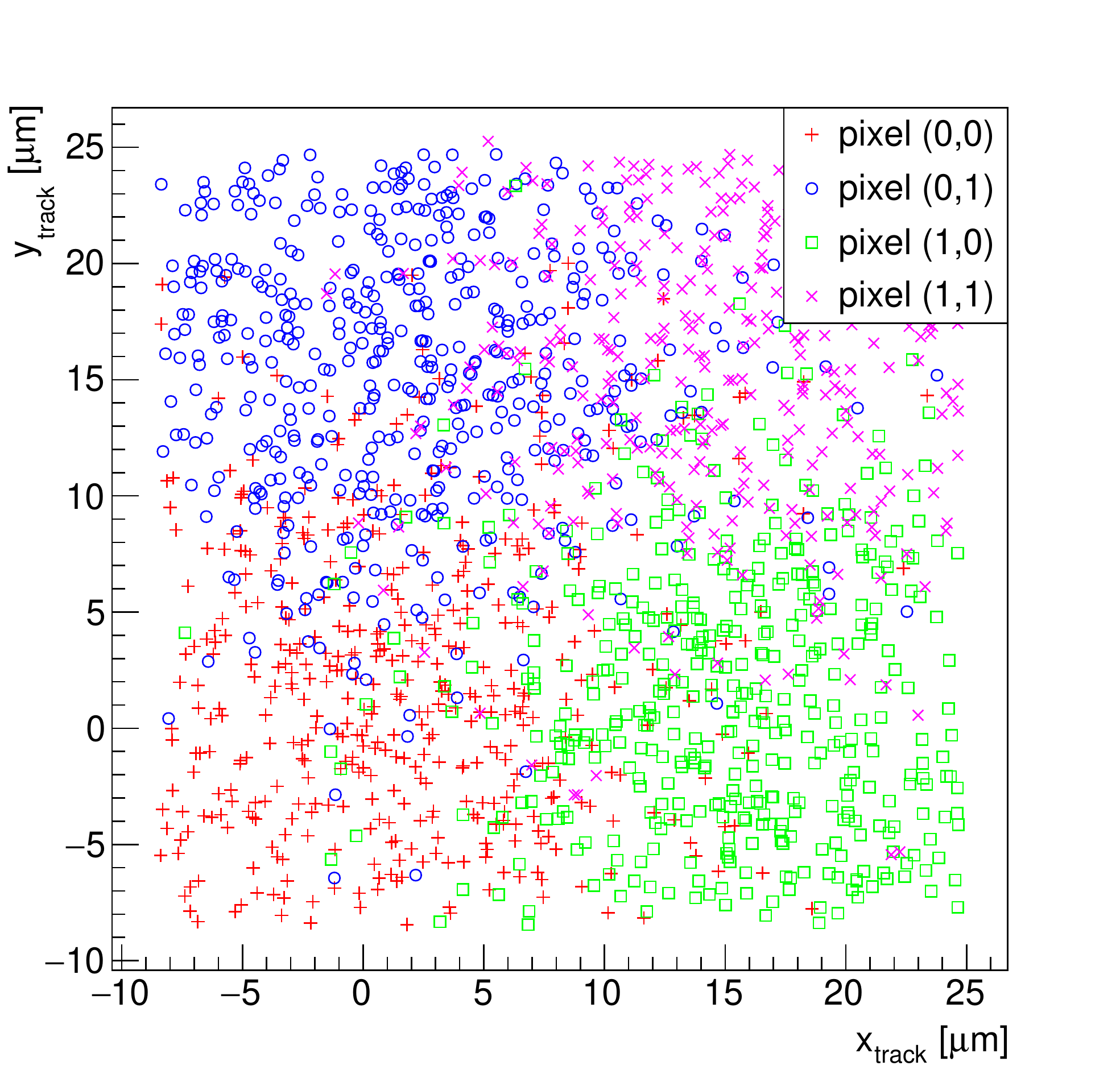}
    \caption{Extrapolated track hit positions on the DESY test chip pixel matrix, at the MAMI test beam. The hits are shown with different symbols and colours depending on which of the four pixels registers a hit.}
    \label{fig::associatedTrackPositions_1}
\end{figure}

Test beam studies have been carried out at the DESY II Test Beam facility~\cite{desyII}, CERN SPS, and at the Mainzer Mikrotron (MAMI) facility~\cite{mamiB}. In a test beam, the device under test is placed in a telescope consisting of multiple reference planes of pixellated sensors. As a beam of particles is sent through the reference planes and the device under test, a reference track can be reconstructed by using hit positions in the telescope. By extrapolating this track, the expected hit position at the device under test can be found. This can then be compared to the sensor output to determine the effect of particles hitting different positions on the sensor. Figure~\ref{fig::associatedTrackPositions_1} shows the extrapolated track position on the DESY test chip at the MAMI test beam, with varying symbols and colours depending on which of the four pixels in the sensor registers a hit. The MAMI facility provides a high-intensity beam of electrons with an energy of 855~MeV.
It can clearly be seen that the four pixels are active in four different regions of the sensor, so it functions in a pixellated manner. Due to the relatively low energy provided by the MAMI facility however, the resolution is dominated by multiple Coulomb scattering in the detector planes. An issue in sensor design after submission has been identified, preventing further studies of pixel characteristics using this test chip. The issue is well understood and will be corrected for future sensor submissions.

\section{Simulation studies}

To understand the sensor behaviour and to optimise the sensor geometry, detailed simulations are performed. Sensor design and submission is slow and costly, so simulations are a useful complement. Technology computer-aided design (TCAD)~\cite{sentaurusTCAD} simulations are used to find the detailed electric field configurations of sensor designs by numerically solving Poisson equations using sensor doping information. Generic doping profiles are used and varied to gain insight into how sensor behaviour is affected by different parameters. These generic profiles are not related to a specific CMOS imaging process, but they are useful for comparative studies. Three main designs are simulated and tested, labelled ``standard layout''~\cite{standardProcess}, ``modified layout''~\cite{modifiedProcess}, and ``n-gap layout''~\cite{ngapProcess}. The standard layout is similar to what is used in the ALPIDE sensor~\cite{alpide}, which is a MAPS used in the ALICE experiment since the latest upgrade (ITS2), developed in a 180\,nm CMOS imaging process. The modified and n-gap layouts are further developments of small collection electrode MAPS, allowing for a higher depletion and improved charge collection characteristics. They were originally developed for a 180\,nm CMOS imaging process, but similar developments have been implemented in a 65\,nm process as well.

The TCAD simulations can also provide information about how a time-resolved charge pulse in the sensor will look and behave, simulating a particle hit. Such simulations take a long time to perform using TCAD however, so in the project the TCAD simulations are combined with Monte Carlo simulations using the Allpix$^2$ framework~\cite{ap2} to generate data with a high statistical significance. Together TCAD and Allpix$^2$ are a powerful combination, making it possible to simulate detailed sensor behaviour and performance with high statistics. This has previously been tested and validated for other MAPS with small collection electrodes~\cite{tcadPlusMC,transientMC}.

\subsection{Technology computer-aided design (TCAD)}

\begin{figure}[tbp]
\centering
    \includegraphics[width=0.45\textwidth, trim={0 0 0 0}, clip]{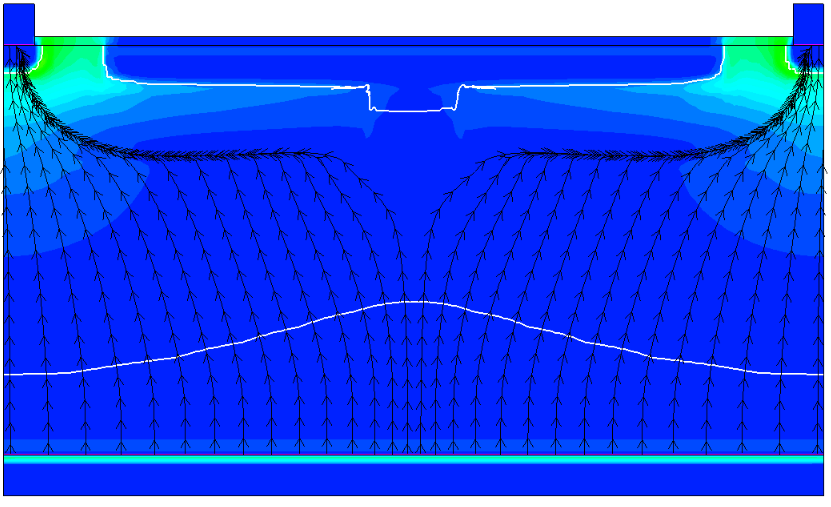}
    \caption{Electric field magnitude output from a TCAD simulation of an n-gap layout. The colour indicates electric field strength, the white lines indicate depletion boundaries, and the black lines are ``streamlines'' for electrons.}
    \label{fig::tcad_ngapFieldExample}
\end{figure}

In TCAD, the doping concentrations and geometries of different parts of the sensor are altered and evaluated, and electric fields are generated. This way, relative measurements can be made between different sensor geometries, even if the doping profiles are generic rather than related to a specific manufacturing process. The simulation studies endeavour to optimise the sensor performance by changing aspects of the sensor design.
Figure~\ref{fig::tcad_ngapFieldExample} shows a typical TCAD simulation electric field magnitude output for an n-gap layout.

In this image, the sensor collection electrodes are located in the upper corners. The sensor is almost fully depleted, and the electric field is strongest closest to the collection electrodes. The black streamlines show the path that an unbound electron in the sensor would follow if it moved solely by drift. They curve towards the collection electrodes, which is an effect brought on by the n-gap layout.

\subsection{Monte Carlo studies using Allpix$^2$}

In Allpix$^2$, the full chain of a detector event can be simulated. This includes energy deposition in a sensor from an incoming particle, charge carrier creation and propagation in the sensor, and signal generation and digitisation. In this project, each pixel in the sensor model contains electric fields and doping concentrations imported from TCAD. Incident particles are generated via an interface to Geant4~\cite{geant4}, and electron-hole pairs are generated as energy is deposited in the sensor along the particle tracks. The electron-hole pairs are propagated to sensor collection electrodes, finally giving the collected charge per incident particle in each pixel. Electronics noise is then added to the collected charge, and a detection threshold is set in the simulations to exclude pixels in which the generated signal does not exceed a given amplitude.
As the simulation is fast, it allows a full Monte Carlo study to be made, using a large number of particles hitting different sensor positions. This makes it possible to account for stochastic fluctuations stemming from the underlying physics processes.

There are three main figures of merit used for the sensor simulation studies: detection efficiency, cluster size, and spatial resolution. The detection efficiency is strongly related to the detection threshold, and should be high across a wide threshold range to simplify design and make the sensor more robust to changes in noise. The cluster size is the number of pixels that register a hit for a single incident particle, and is thus a measure of the charge sharing taking place. This depends on the position of the incident particle in the sensor, but with a large data sample a mean value can be found. The spatial resolution indicates how well the sensor can reconstruct the incident particle position. In the simulations, this is extracted by comparing the Monte Carlo truth particle position to the reconstructed position, which is taken as the charge-weighted mean position of a cluster of pixels. The spatial resolution is taken to be the root mean square of the central 99.73\% of the resulting distribution (corresponding to the central $\pm 3 \sigma$ of a Gaussian distribution).

\begin{figure}[tbp]
\centering
    \includegraphics[width=0.45\textwidth, trim={0 0 12mm 12mm}, clip]{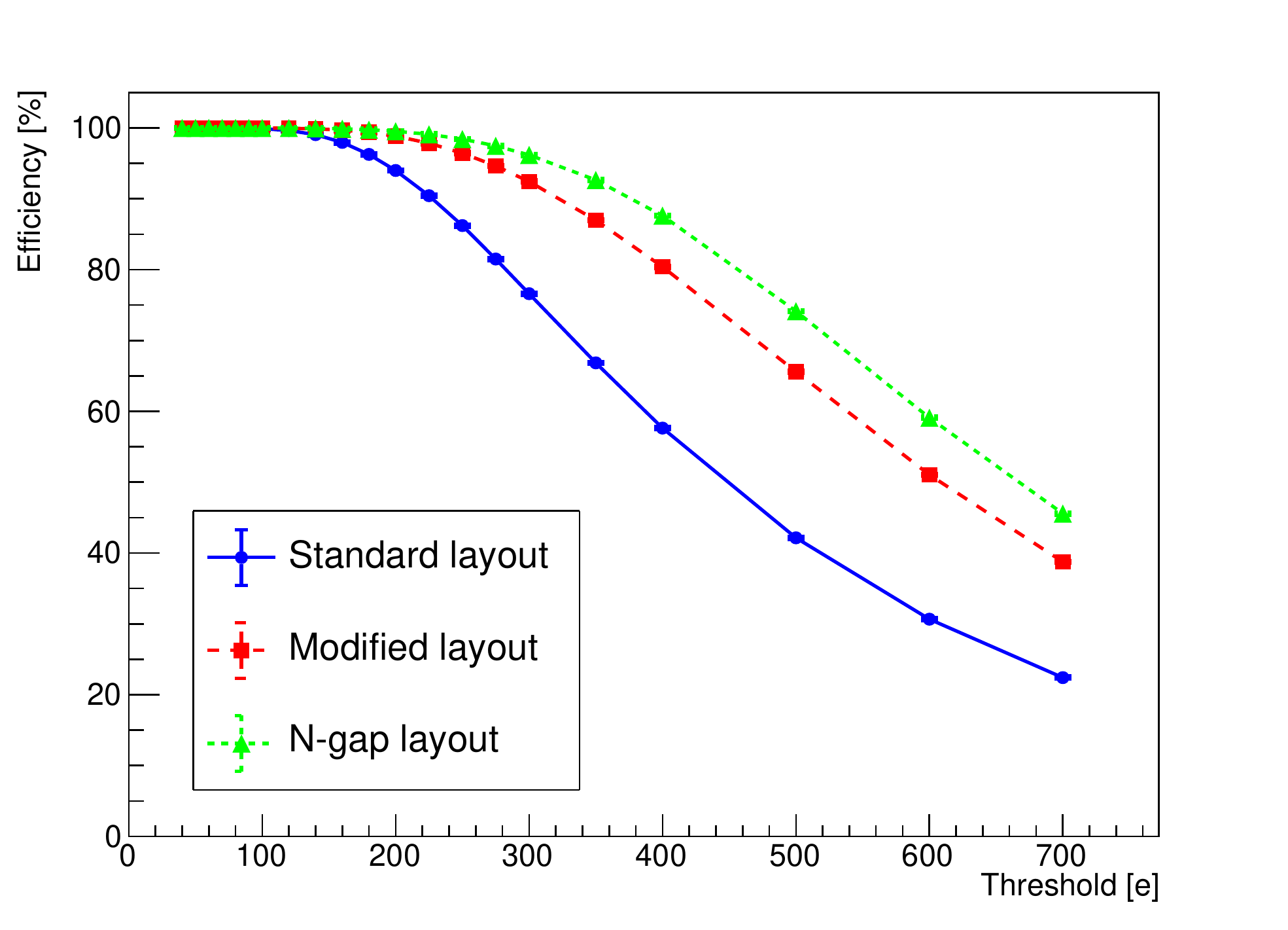}
    \caption{Mean efficiency versus detection threshold for three different pixel layouts.}
    \label{fig::ap2_processComp_efficiency_1}
\end{figure}

In the simulations presented here, a beam of 5~GeV electrons is simulated, incident head-on on a single pixellated sensor with a pixel size of $20 \times 20$\,\textmu m$^2$. For each investigated configuration 500~000 single-electron events are simulated, which means that the statistical errors in the results are small. The detection threshold is varied, and the figures of merit plotted versus the threshold value. Figure~\ref{fig::ap2_processComp_efficiency_1} shows simulation results for the mean efficiency across a sensor. The figure shows the standard, modified, and n-gap layouts with different line styles.

From these results, it can be seen that the modifications to the sensor layout are beneficial as they increase the efficient operating margin of the sensor. A sensor with the n-gap layout maintains high efficiency for a larger range of threshold values than a sensor in the standard layout. The current baseline threshold for the DESY sensor developments is 200 electrons.

\begin{figure}[tbp]
\centering
    \includegraphics[width=0.45\textwidth, trim={0 0 12mm 12mm}, clip]{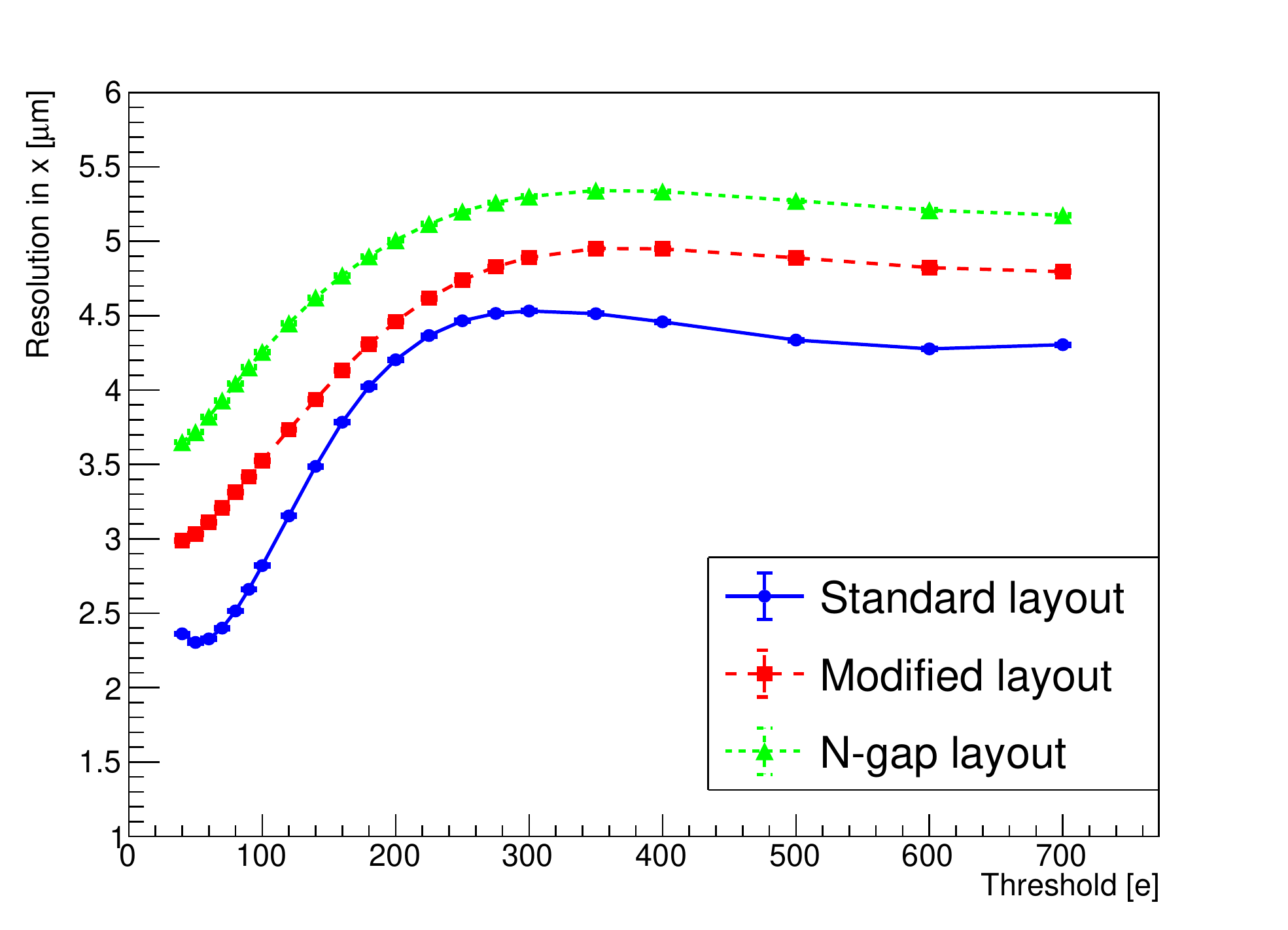}
    \caption{Spatial resolution versus detection threshold for three different pixel layouts.}
    \label{fig::ap2_processComp_residualX_1}
\end{figure}

Figure~\ref{fig::ap2_processComp_residualX_1} shows the sensor spatial resolution in the x direction for three different layouts. As the simulated pixels are square and symmetric, the resolution is identical in the y direction.
A lower value means a better resolution, so it can be seen that resolution deteriorates as threshold increases, and that the standard layout has the best resolution of the three. This is due to the reconstructed position being more accurate when the cluster size is larger, due to charge-weighted interpolation between more pixels. As the standard layout has a larger undepleted region at pixel edges, charges there move primarily by diffusion and spread more to neighbouring pixels, increasing the cluster size. The n-gap layout is made to funnel charge more towards the collection electrodes, and thus has less charge sharing and a worse spatial resolution.

Other simulation studies have also been performed, for example comparing different pixel sizes, different sensor bias voltage configurations, and different geometries. There is significant progress in understanding the impact of different parameters, and the conceptual design of a new sensor is being converged on.

\section{Conclusions and outlook}

The DESY prototype chip has been tested thoroughly, both in labs and at test beams. The sensor is functional and tests can be performed on the sensor electronics.
In collaboration with colleagues at CERN, tests are starting on an analogue pixel test structure which will be used to compare different sensor designs and see if simulation results match reality. There are two test beam campaigns planned using this test structure; one at the MAMI facility in April 2022, and one at the DESY II Test Beam facility in June 2022.

The simulations show that both efficiency and spatial resolution benefit from a low detection threshold. The threshold should thus be kept as low as possible, but what is possible depends heavily on the electronics design and noise. The standard layout has a better spatial resolution than the other tested layouts, but a smaller operating margin in terms of efficiency. A balance thus needs to be reached for the final sensor design.
Simulations using generic doping profiles provide valuable results for use in sensor optimisation, and help in understanding sensor behaviour. The simulations are ongoing, and by using the method of combining TCAD simulations and Monte Carlo simulations high-statistics data of different situations can be produced relatively quickly. As data from test chips become available, simulations will be validated against those.

The next sensor submission will contain the next iteration of a DESY prototype test chip in a 65\,nm CMOS imaging process. For this chip the electronics design will be updated, and the chip will contain a larger pixel matrix. The new sensor is expected to be available at DESY at the end of 2022.

\section*{Acknowledgements}

The Tangerine project receives funding from the Helmholtz Innovation Pool, 2021 - 2023. Measurements leading to some of the presented results have been performed at the Test Beam Facility at DESY Hamburg (Germany), a member of the Helmholtz Association (HGF).

The authors wish to express their gratitude to the Institut für Kernphysik at the Johannes Gutenberg-University in Mainz, and the MAMI test beam and its operating crew.
We are also grateful for the help and support of the ALICE ITS3 project, and the CERN SPS test beam facility.

\bibliography{mybibfile}

\end{document}